\pdfoutput=1

\documentclass[twocolumn,showpacs,preprintnumbers,amsmath,amssymb]{revtex4-1}
\usepackage{dcolumn}
\usepackage{bm}

\usepackage[dvips]{color}
\usepackage{ulem}
\usepackage[pdftex]{graphicx}
\usepackage{epstopdf} 
\usepackage{amsmath}

\begin{document}

\preprint{APS/unknown}

\title{Exchange anisotropy pinning of a standing spin wave mode}

\author{R.  Magaraggia}
\author{M. Kostylev}%
\author {K. Kennewell}%
\author {R. L. Stamps}%
\affiliation{School of Physics, University of Western Australia, 35 Stirling Highway, Crawley, Western Australia 6009, Australia }

\author{M. Ali}
\author{D. Greig}
\author{B. J. Hickey}
\author{C. H. Marrows}
\affiliation{School of Physics and Astronomy, E. C. Stoner Laboratory, University of Leeds, Leeds LS2 9JT, United Kingdom }%

\date{\today}

\begin{abstract}
Standing spin waves in a thin film are used as sensitive probes of interface pinning induced by an antiferromagnet through exchange anisotropy.  Using coplanar waveguide ferromagnetic resonance, pinning of the lowest energy spin wave thickness mode in $\text{Ni}_{80}\text{Fe}_{20}$/$\text{Ir}_{25}\text{Mn}_{75}$ exchange biased bilayers was studied for a range of IrMn thicknesses.  We show that pinning of the standing mode can be used to amplify, relative to the fundamental resonance, frequency shifts associated with exchange bias. The shifts provide a unique `fingerprint' of the exchange bias and can be interpreted in terms of an effective ferromagnetic film thickness and ferromagnet/antiferromagnet interface anisotropy. Thermal effects are studied for ultra-thin antiferromagnetic $\text{Ir}_{25}\text{Mn}_{75}$ thicknesses, and the onset of bias is correlated with changes in the pinning fields.  The pinning strength magnitude is found to grow with cooling of the sample, while the effective ferromagnetic film thickness simultaneously decreases. These results suggest that exchange bias involves some deformation of magnetic order in the interface region.
\end{abstract}

\maketitle

\section{Introduction}
Exchange bias is an effect which has consequences for the bulk of a ferromagnet as exhibited by hysteresis loop offset.  However its bulk effects arise from coupling processes across a ferromagnetic/antiferromagnetic interface\cite{Ex_Bias_Review_Robert,Nogués1999203}.  Directly probing these types of buried interfaces to gain information on coupling is quite challenging.
Ferromagnetic resonance (FMR) is a powerful tool for studying magnetic parameters in ferromagnetic structures through frequency shifts of the fundamental resonance mode. It is possible to also use FMR to detect standing spin waves which provide, at least in principle, information about surfaces and buried interfaces\cite{PhysRevB.38.6847,PhysRevB.58.8605,PhysRev.110.1295}.  In this paper standing spin waves (also referred to as ``thickness modes") are used to probe interface properties due to exchange anisotropies in exchange biased bilayers. We show that a useful measure for characterising exchange bias can be obtained from these modes, and this measure can provide unique information about magnetic ordering in the interface region.

Nearly all studies of ferromagnetic resonance and spinwaves in exchange biased structures have, to date,  made use exclusively of the fundamental resonance or zone center spinwaves\cite{Ex_Bias_FMR_Rotation,PhysRevB.70.094420,PhysRevB.58.8605}.  The frequencies of these excitations are governed primarily by local magnetocrystalline and shape anisotropies, magnetization, and applied field. The resonance conditions for a ferromagnetic thin film with no intrinsic anisotropies, and magnetised in plane, is given by\cite{Kittel_Formula}:

		\begin{align}
	&(\frac{\omega}{\gamma})^{2} = (H_{f}(\theta) + D k_{y}^{2}(\theta))(H_{f}(\theta) + \mu_{0}M_{s} + D k_{y}^{2}(\theta))&
		\label{equ:kittel}
		\end{align}
		
\noindent The spin wave frequency is $\omega$, $\gamma$ is the gyromagnetic ratio, $M_s$ is the saturation magnetisation, $H_f$ is the field applied to cause resonance, and $\theta$ is the direction of the applied field relative to the cooling field direction.  A fixed spin wave frequency is assumed and $\theta$ is varied, so that $H_f$ becomes the experimentally meaured quantity.  The wavevector component in the direction normal to the film plane is $k_y$. The $\mu_{0}M_{s}$ term originates from dynamic demagnetisation fields in thin film geometry, and $D=\frac{2A}{M_{s}}$ is the exchange coupling strength.  In traditional treatments of FMR as applied to exchange-bias the fundamental FMR mode corresponds to $k=0$. Effective fields originating at the interface with the antiferromagnet are then, as far as the FMR response is concerned, averaged over the ferromagnetic film thickness and are seen as an effective anisotropy field. In a resonance experiment using a fixed frequency, these effective fields appear in the measured value of $H_f$, the applied field for which resonant absorption is observed. It is important to note that the frequency shifts of the FMR associated with exchange bias do not contain direct information about the interface region per se. Questions concerning the penetration depth of the  interface fields, or asymmetries associated with different boundaries, can only be addressed indirectly by varying film thicknesses within a series of samples. A disadvantage of this approach is that samples can vary substantially, even within the same series due to details of growth processes\cite{Berkowitz1999552,Nogués1999203}.

The FMR mode averages local interface fields laterally because it is a long wavelength excitation, though in reality it does experience deformation due to the interfacial pinning. In some cases, short wavelength spin waves can be observed with conventional resonance techniques as standing wave thickness modes confined by film geometry.  It is access to these modes which allows a measure of interface pinning.  Recently we have shown theoretically and experimentally that broadband FMR driving techniques that make use of stripline or coplanar waveguides can couple effectively to thickness modes in metallic multilayers\cite{kostylev-2008,screening_unpublished}.  These thickness modes have some discrete wavevector $k_{y}(\theta)$, and therefore involve contributions from exchange.  Hereafter these modes are referred to as ``FEX modes". These will each have different allowed wavevectors confined in the $y$ direction, as determined by surface pinning. As such, the frequencies of the FEX modes include contributions from exchange, and are sensitive to surfaces and interfaces.  The lower symmetry at film boundaries can give rise to local anisotropy fields, and interfaces between different magnetic layers can support exchange coupling. In these cases, spin wave oscillations may be pinned at one or more boundaries of a ferromagnetic film. Pinning of this type is accompanied by contributions through exchange energies, and can result in substantial frequency shifts \cite{PhysRevB.54.4159}.

A simple means of analysing frequencies obtained for thickness modes was suggested long ago by Rado and Weertman\cite{Spin_Wave_pinning,PhysRev.110.1295,PhysRevB.19.4575}. In this approach, surface anisotropies are assumed, which then dictate the boundary conditions for FEX modes in thin film geometries.  It should be noted that the FMR mode will also be affected and given a non-zero wavevector resulting from surface pinning.  If we associate a surface energy\cite{PHYS.REV.B.72.014463.(2005)} of the following form with the exchange biased interface:

	\begin{eqnarray}
	E_{SA} = p\cdot\:M_{s}
	\label{equ:pinningenergy}
	\end{eqnarray}
	
\noindent We can then calculate allowed spin wave wavevectors as a result of pinning. In this equation $M_{s}$ is the saturation magnetisaion and $p$ is the pinning parameter which acts parallel to the applied field.  As demonstrated in \cite{Spin_Wave_pinning}, if one starts with the Landau-Lifshitz equation and integrates over an infinitesimal volume region across the interface, the following is obtained:

		\begin{eqnarray}
	  (\frac{2A}{M_{s}}) \boldsymbol{M}\times \frac{\partial\boldsymbol{M}}{\partial n} + \boldsymbol{T_{surf}}=0
		\label{equ:pinninglandau}
		\end{eqnarray}

\noindent Here \textbf{M} represents the total magnetisation, $n$ is the direction normal to the interface and $\boldsymbol{T_{surf}}$ is the interface torque. Using Eq.(\ref{equ:pinningenergy}), we have:

		\begin{eqnarray}
	  \boldsymbol{T_{surf}}= -\boldsymbol{M} \times \nabla_{M} E_{SA} = -\boldsymbol{M} \times p
		\label{equ:pinningefffie}
		\end{eqnarray}

\noindent We approximate the exchange biased interface by supposing the pinning to come entirely from one of the boundaries, hence introducing an asymmetry into the model.
After solving Eq.(\ref{equ:pinninglandau}) in combination with Eq.(\ref{equ:pinningefffie}), the relationship between these surface anisotropies and $k_y$ for $H_{f}$ applied at an angle $\theta$ to the easy axis is:

		\begin{eqnarray}
	&p(\theta) = (\frac{2A}{M_{s}}) (\frac{-k_{y}(\theta)}{\cot(k_{y}(\theta) t_{eff})})
		\label{equ:pinningk}
		\end{eqnarray}
					
\noindent It is important to note that $t_{eff}$ is the magnetic thickness of the ferromagnet, as opposed to the structural thickness (which may be different)\cite{bruck:126402,PhysRevLett.95.047201}.  This difference may be caused by deviations away from uniform ferromagnetic order near the interface due to local pinning fields.

The remainder of the paper is organized as follows. First, preparation of, and magnetization measurements from, exchange biased $\text{Ni}_{80}\text{Fi}_{20}$/$\text{Ir}_{25}\text{Mn}_{75}$ are discussed. Next we present results from coplanar FMR studies of the fundamental and first thickness modes for these structures, and discuss their interpretation in terms of the pinning parameter $p$ and effective thickness $t_{eff}$.

\section{Sample Preparation and Characterization}

Magnetic bi-layer specimens consisting of $\text{Ta(50 \AA)/ Ni}_{80}\text{Fe}_{20}\text{ (605 \AA)/ Ir}_{25}\text{Mn}_{75}\text{ ( t}_{AF}\text{ \AA)/ Ta(50 \AA)}$ were sequentially deposited onto Si(001) substrates by dc-magnetron sputtering at an argon working pressure of 2.5 mTorr to minimise growth variations.  A nanometer layer of native oxide on the silicon surface created conditions for polycrystalline growth. Typical deposition rates were 2$-–$2.5 \AA  $\text{s}^{-1}$, which were determined by measuring the thickness of calibration films by low-angle x-ray reflectometry. The base pressure prior to the deposition was of the order of $1\times10^{-8}$ Torr and the samples were deposited at ambient temperature. An in-plane forming field of 200 Oe was applied during the growth to induce a macroscopic uniaxial anisotropy in the NiFe (Py) layer in a defined direction. The thickness of the IrMn layer, $\text{t}_{AF}$, for this study was varied from 0 to 60 \AA \ which is also the region where the onset of biasing appears at room temperature for such systems\cite{PhysRevB.67.172405}.  The samples were cut into $\text{10mm}\times\text{10mm}$ squares.

\begin{figure}[ht]
		\includegraphics[width=9cm]{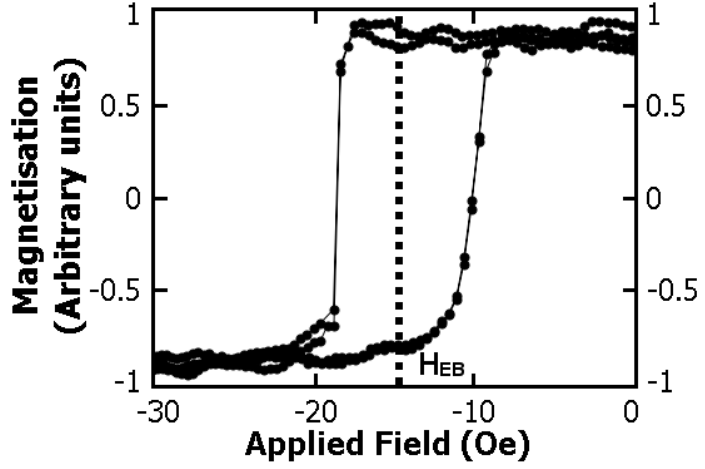}
		\caption{\label{fig:moke} Shown above is a sample of data taken with a MOKE magnetometry setup focused onto the NiFe(60.5nm)/IrMn(6nm) sample. The vertical axis uses arbitrary units and represents the average magnetisation over the laser spot focused onto the sample.  The horizontal axis displays field applied across the sample in units of Oersteds.  Also the exchange bias shifting of the loop is shown by the dotted line and denoted by H$_{EB}$.}
		\end{figure}

Film thickness was accurately characterised with a Siemens two-circle diffractometer, to within $\pm$6 \AA.  In-plane and out-of-plane FMR magnetometry was used to extract $\mu_{0}M_{s}$, which could consistently be used in further FMR data analysis.
In-plane FMR magnetometry along the easy axis of a Py sample with no IrMn revealed a saturation magnetisation $\mu_{0}M_{s}$ of 0.80$\pm$0.05T, a gyromagnetic ratio $\gamma$ of 2.8$\times \text{10}^{10}$Hz$\text{T}^{-1}$ and in plane bulk anisotropy fields of 0.0002T$\pm$0.0005T.
Further magnetometry was performed using the magnetoopical Kerr effect (MOKE). A $635$nm diode laser, rated at 5mW, was used to illuminate the sample.  A differential amplifier was used to analyse polarisation rotation. Example results are shown in Fig. \ref{fig:moke}.

			\begin{figure}[h!tb]

		\includegraphics[width=8.5cm]{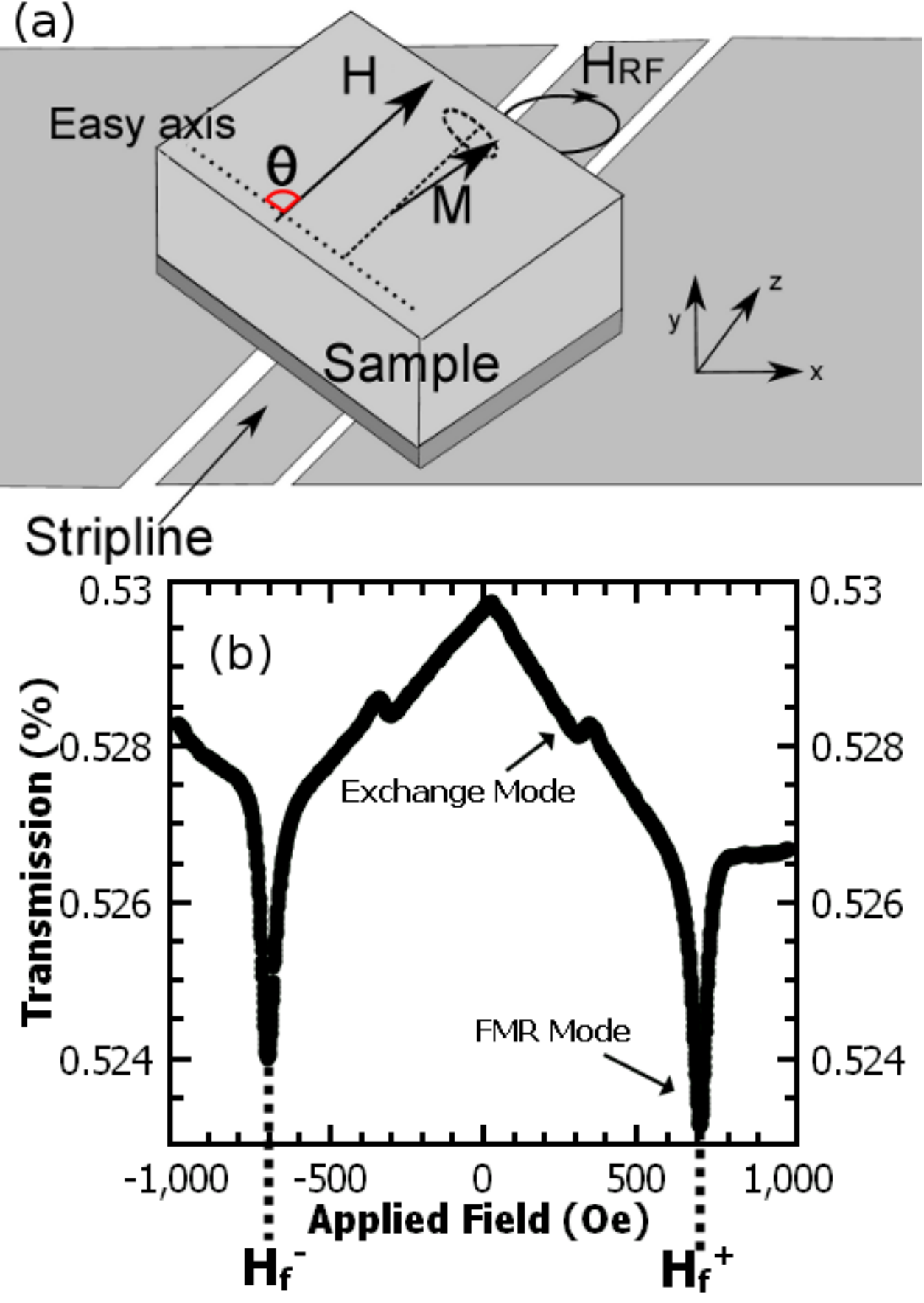}
		\caption{\label{fig:film} Panel (a) shows the experimental geometry, with the sample placed on top of the coplanar stripline.  $H$ refers to the applied field direction at some angle $\theta$, $M$ refers to the magnetisation direction and $H_{RF}$ demonstrates the microwave rf field generated by the waveguide.  The sample is rotated in place in order to change the direction of $H$ with respect to the sample's easy axis. Panel (b) shows microwave transmission as a function of static applied field for the 0nm IrMn sample. The values $H_{f}\pm$ correspond to applied resonant fields in antiparallel directions for + and - respectively.  Microwave absorbtions are seen which correspond to the fundamental mode (FMR) and the first exchange mode (FEX).  The microwave excitation frequency $\omega$ used was 7GHz. }

		\end{figure}
				
As demonstrated in Fig. \ref{fig:moke}, the samples saturate magnetically above 20Oe.  The loops are non-symmetric about a non-zero field with a small coercivity, and compare well with what has been found in similar studies\cite{PhysRevB.70.094420,PhysRevB.63.174419}.  The bias field as measured with FMR is defined as $H_{EB} = \frac{H_{f^{+}} - H_{f^{-}}}{2}$, shown in Fig.\ref{fig:film}, where $H_{f^{+}}$ corresponds to $H_{f}(0)$ in eq.\ref{equ:kittel}, and $H_{f^{-}}$ corresponds to $H_{f}(\pi)$. The coercivity increases with increasing IrMn thickness, with a maximum in the region of thicknesses where there exists little exchange biasing.

\section{Resonance Measurements and Interpretation}

A 20GHz Vector Network Analyser was used to excite and detect FMR and FEX modes of the samples.  The coplanar stripline (0.3mm wide) which is coupled to 50$\Omega$ axial cables, excites the sample with microwaves in the 2-9 GHz regime.  Example results are shown in Fig.\ref{fig:film}.  We choose a particular excitation frequency $\omega$ and sweep the applied magnetic field $H$ (usually between 0 and 600 Oe), in a particular direction until microwave power is absorbed strongly by the sample, indicating a standing spin wave is on resonance.  This procedure is repeated for the samples' easy axis aligned along different directions with respect to the applied field, denoted by $\theta$.  A field sweep was chosen rather then a frequency sweep, as a field sweep avoids the problems of variable microwave frequency attenuation in the waveguides with varying $\omega$ and shows the magnetic response of the sample as opposed to both magnetic and electric response.
		

An example of FMR and FEX resonances, at a driving frequency of 7 GHz, is shown in Fig.\ref{fig:film}(b). A number of factors determine the observed amplitudes of FMR and FEX modes in coplanar geometries \cite{kostylev-2009,PhysRev.97.1558,PhysRevLett.90.227601,PhysRev.118.658}, in particular a combination of surface pinning and eddy current induced inhomogeneity in the driving microwave field.  The FEX absorpton amplitude is approximately 23 times less than that of the FMR mode as measured at 7GHz.  The linewidths of the modes at 7GHz are $\Delta f_{FMR}$=49Oe and $\Delta f_{FEX}$=25Oe respectively. It should be noted that the FMR mode has a Lorentzian like absorption shape, but the FEX mode does not, so the linewidths may not be directly comparable.

\begin{figure}[ht]
		\includegraphics[width=8cm]{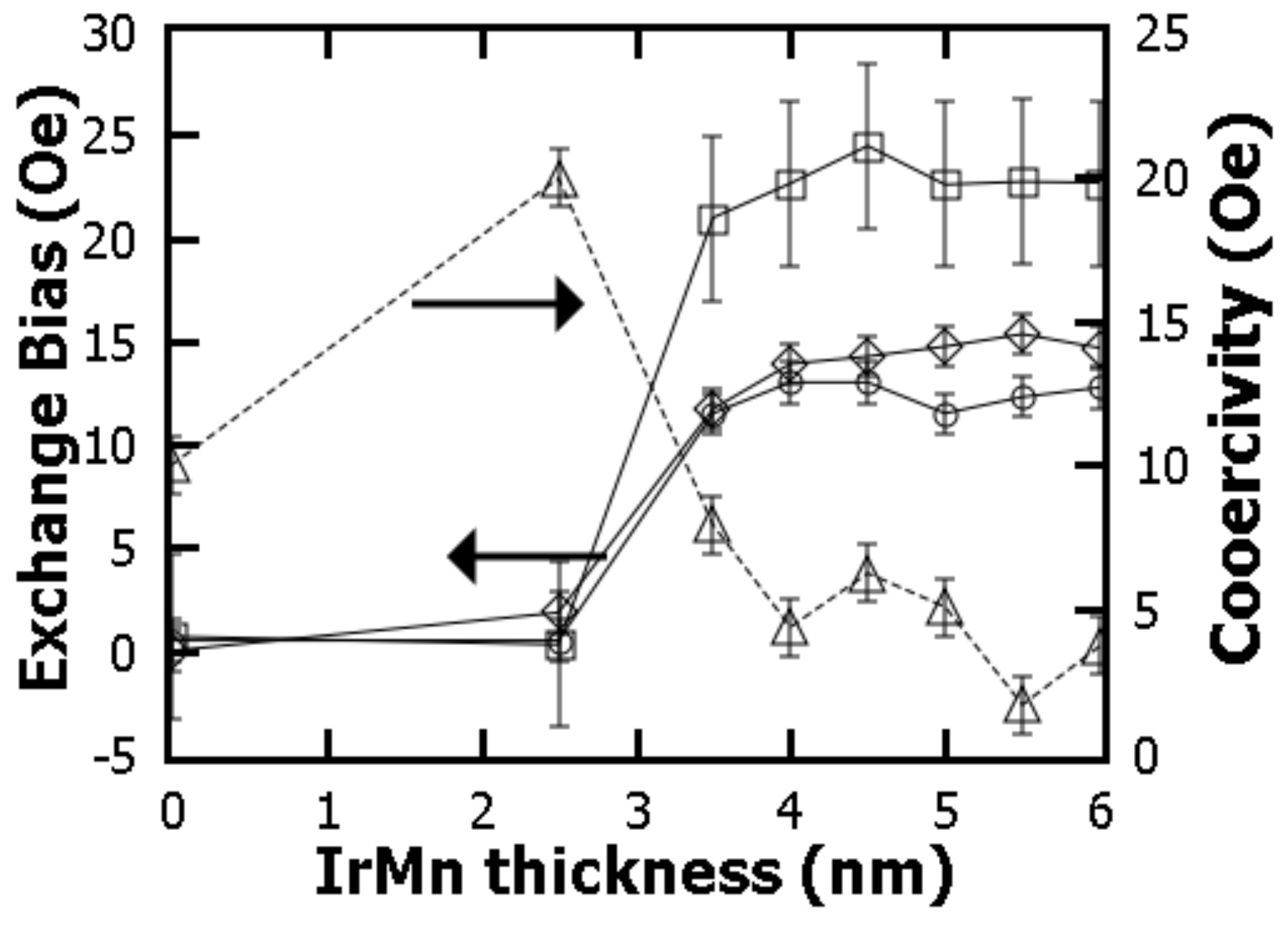}
		\caption{\label{fig:bias} Shown is the exchange bias as measured from the FMR mode (empty circle solid line), MOKE (empty diamond solid line) and FEX mode (empty square solid line), as a function of IrMn film thickness.  The NiFe layer thickness is always 60.5nm.  For comparison the cooercivity as measured with MOKE is shown (hollow triangle dashed line)}
		\end{figure}

The bias determined from FMR and FEX are shown in comparison to the bias determined from MOKE data in Fig.\ref{fig:bias}. Unidirectional exchange anisotropies are present at room temperature only for a certain critical thickness $>$ 2.5nm of IrMn as shown in Fig.\ref{fig:bias}. 

\begin{figure}[ht]
		\includegraphics[width=7cm]{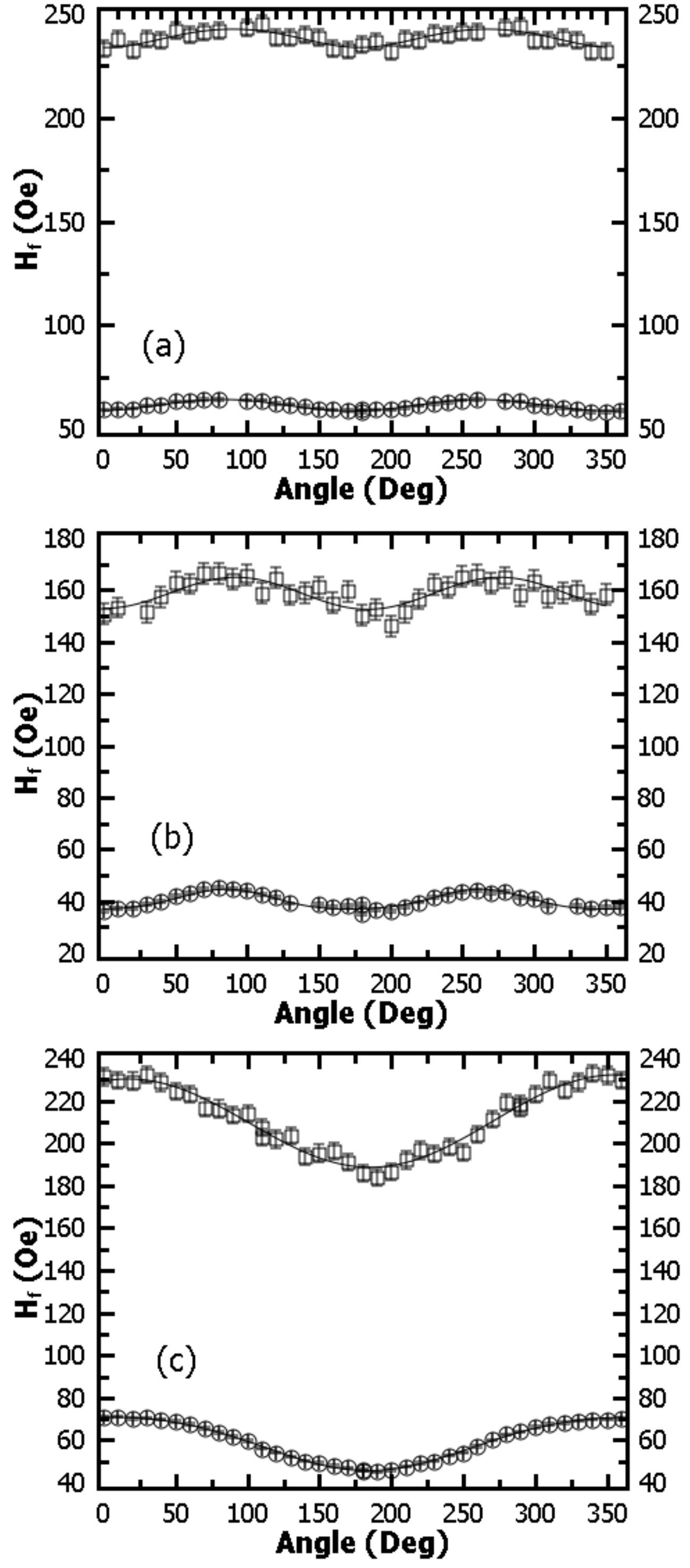}
		\caption{\label{fig:rotresonance} Shown are the resonant fields $H_{f}$ for the FMR (empty circle solid line)) and FEX (empty square dashed line) standing spin wave modes at different applied field angles with respect to the easy axis ($\theta$).  The solid lines show fits to the data using $cos(\theta)$ and $cos(2\theta)$ components.  Presented is the resonance data for different IrMn thickness capping layers a) IrMn=0nm, b) IrMn=2.5nm, c)IrMn=6nm.}
		\end{figure}

For thicknesses above this value, the MOKE and FEX results indicate a non-monotonic behavior of the bias with respect to IrMn thickness begining at 4 nm. We have at present no explanation for this, though this could be due to sample to sample variation. It is possible that it may have other origins, as such behaviour in similar systems has been noted and explained via the domain-state model\cite{PhysRevB.68.214420}.  Most significantly, the pinning field is unidirectional. This is fully consistent with exchange bias as an interface effect. The bias acts as an effective volume unidirectional anisotropy when averaged by the FMR mode, and appears as a superposition with other volume anisotropies. This superposition can be seen most clearly by measuring bias at different orientations of the applied field relative to the bias field direction. Example results for the 2.5 and 6 nm thick IrMn samples are shown in Fig. \ref{fig:rotresonance}. Results for FMR and FEX peaks are shown as function of angle, demonstrating that both modes contain equal contributions from a uniaxial anisotropy, whereas the modes are affected differently by the exchange bias.

The results shown in Fig.\ref{fig:rotresonance} illustrate the magnitude of exchange bias as measured by the FMR and FEX modes. The difference in magnitude can be understood through pinning effects on the frequency of the FEX modes. The FEX modes contain greater exchange energy than the FMR because of their shorter wavelengths, and pinning acts to effectively change the wavelength of an FEX mode. In this way, pinning by exchange bias is an amplification of exchange anisotropy by affecting directly the exchange energy contribution to an FEX mode.
This is demonstrated explicitly in Eq.(\ref{equ:kittel}), where the exchange-related effective anisotropy field $Dk^{2}_{y}$ scales as the square of the wavenumber $k_{y}$.  Therefore one should expect different strengths of effective anisotropy from the FMR and FEX modes.  Indeed, such differences are seen in Fig.\ref{fig:rotresonance} for these two modes, confirming the interface origins of the anisotropy fields in this exchange biased system.

\begin{figure}[ht]
		\includegraphics[width=7cm]{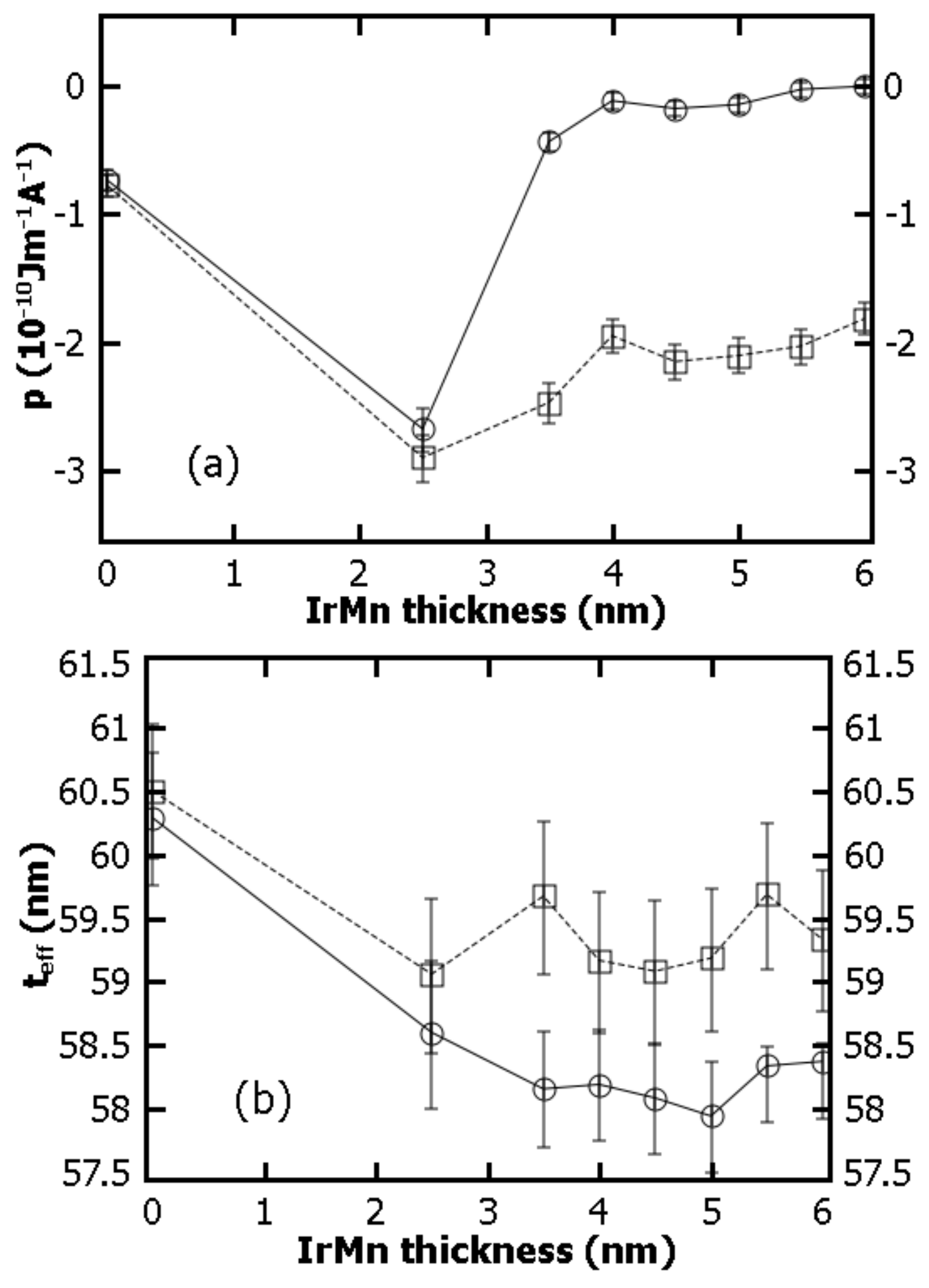}
		\caption{\label{fig:thickpinning} a) The calculated strengths of pinning p along the bias direction (empty circle solid line) and against the bias direction (empty square dashed line).\\
		b) The corresponding effective magnetic thickness $t_{eff}$ of the NiFe along the bias direction (empty circle solid line) and against the bias direction (empty square dashed line).}
		\end{figure}

 Pinning factors $p$, calculated according to Eq.(\ref{equ:pinningk}) as a function of IrMn thickness are shown in Fig.\ref{fig:thickpinning} for data taken at room temperature. Interface anisotropy calculated for the applied field along  $\theta = 0$ is denoted $p_{+}$, and represents the situation there the applied field is antiparallel to the bias field direction.  Conversely, $p_{-}$ is the pinning calculated for the field applied along the bias direction $\theta = \pi$. In these calculations, we have used material parameters determined experimentally as above.  The exchange coupling strength D=$1.3693\times10^{-17}J A^{-1}$ was chosen such that an effective thickness of 60.5nm was extracted from the monolayer permalloy film. Error bars in Fig.\ref{fig:thickpinning} were estimated by incorporating experimental field uncertainties.  We consider $p$ as the more fundamental quantity then exchange bias field.  Pinning will act with the same strength on both modes, but the wavelength of each mode will be distorted to a different degree.  Importantly, in our fittings we have the condition that $p$ should have the same value for all observed modes.  We find this condition cannot be satisfied unless some value is modified for one of the physical parameters in Eq.\ref{equ:pinningk}. The derivation of Eq.\ref{equ:pinningk} and previous works \cite{bruck:126402,PhysRevLett.95.047201} suggest that the suitable parameter is the thickness of the ferromagnetic layer.  Therefore the second parameter extracted from the fits is the effective thickness of the ferromagnetic layer.  As previously mentioned, the difference between $t_{eff}$ and the structural thickness of the ferromagnet might be related to deviation from uniform ferromagnetic order close to the interface.

 The dependence of $p$ on IrMn thickness shows a curious peak for the 4 nm thick film, but otherwise is a nearly linear function of $t_{AF}$ above 2.5 nm.
In addition to an interface pinning, we also simultaneously extract an effective magnetic thickness $t_{eff}$ from the data.  The greatest change of $t_{eff}$ with in-plane field direction appears for $t_{AF}$ between 5 and 5.5nm, a range in which the largest degree of exchange bias is observed with MOKE but not FMR.

	\begin{figure}[h!t]
		\includegraphics[width=7cm]{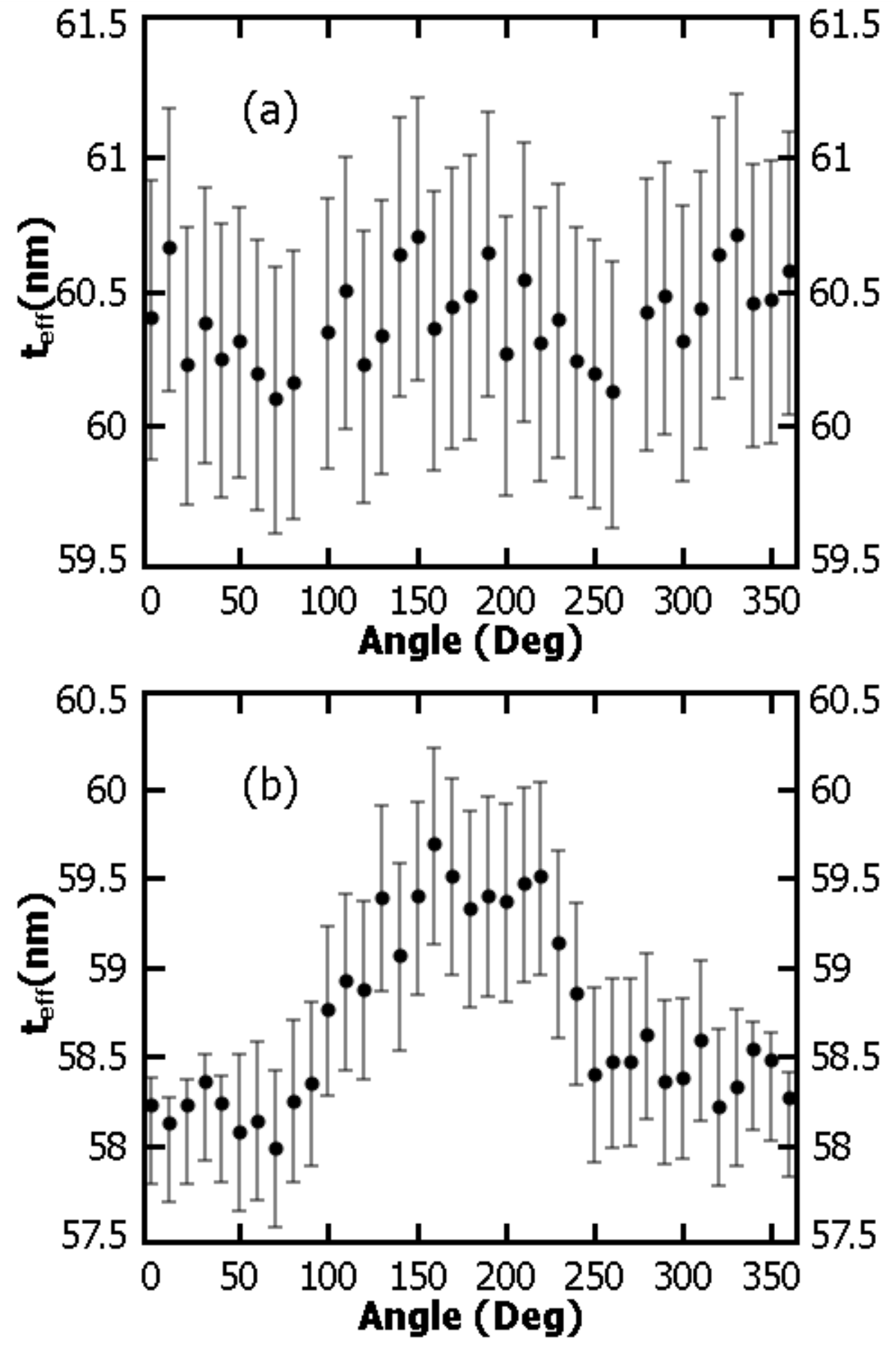}
		\caption{\label{fig:effthick} The effective magnetic thickness of NiFe as a function of $\theta$ with respect to the easy axis for a) 0nm IrMn film,  b) 6nm IrMn film.}
		\end{figure}
		
 Like $p$, the effective thickness varies as a function of applied field direction.  The IrMn free permalloy layer (Fig.\ref{fig:effthick}a) does not show any significant variation of $t_{eff}$ with $\theta$ with the implication that no significant micromagnetic configurational changes take place when aligning the magnetisation along different anisotropy directions.  This is in sharp contrast to the 6nm IrMn film (Fig.\ref{fig:effthick}b), which does display a rougly 1 nm thickness variation of $t_{eff}$ over the angular range $0$ to $180^o$.

 An interpretation of effective magnetic film thickness is difficult as it does not allow identification of specific micromagnetic structures across the interface region.  Nevertheless, it does not seem unreasonable that $t_{eff}$ provides some measure of the size over which magnetization in the interface region contributes to pinning, perhaps through local modification of the magnetic order\cite{PhysRevLett.95.047201,PhysRevLett.91.017203}.

\begin{figure}[h!t]
		\includegraphics[width=7cm]{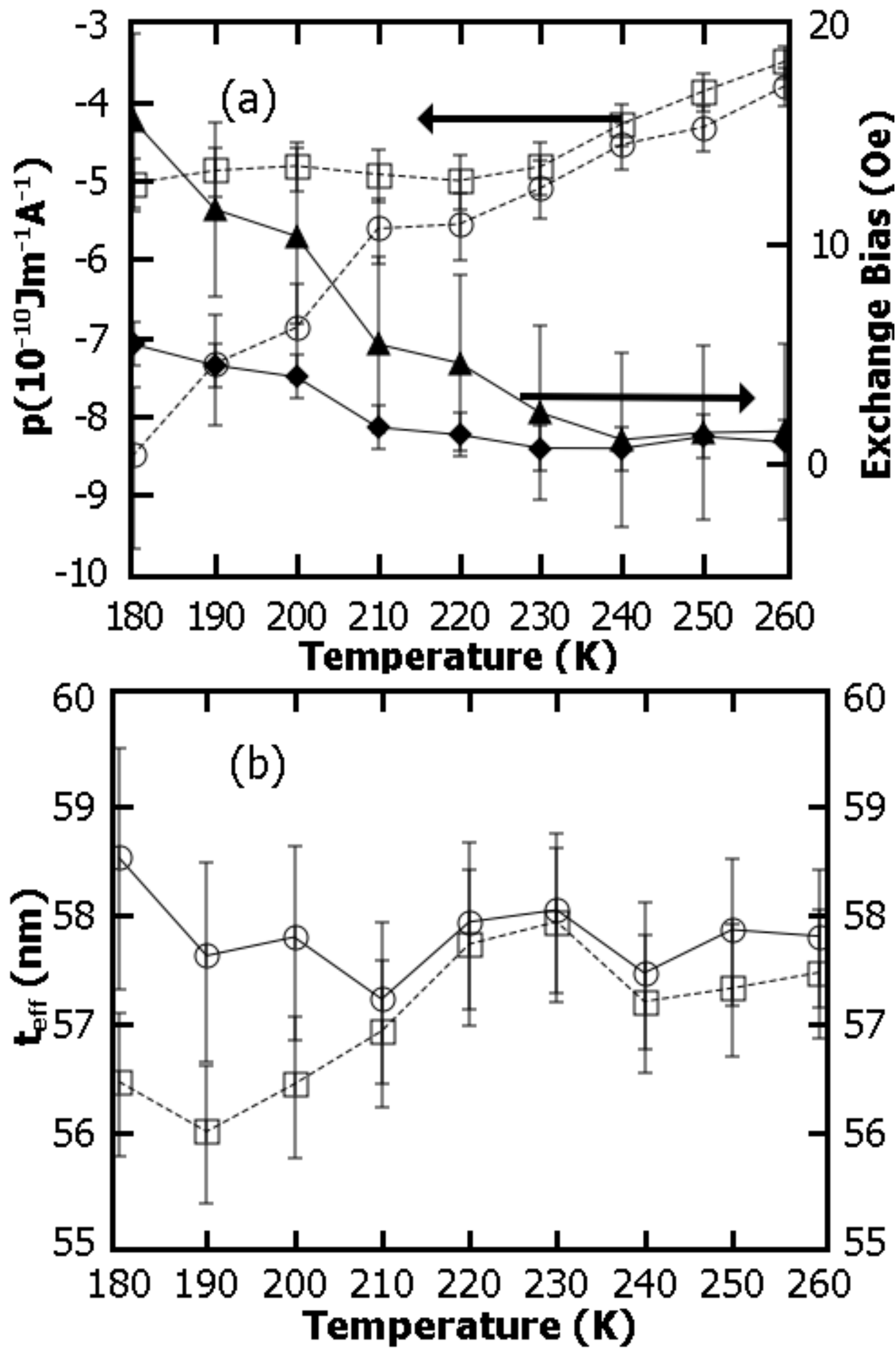}
		\caption{\label{fig:temppinning} a) This figure illustrates the calculated strengths of pinning p along the bias direction (empty circle dashed) and against the bias direction (empty square dashed line) for the IrMn 2.5nm film cooled to the temperature indicated on the horizontal axis, in a 40Oe field.  Also the complementary information on the exchange bias shift for the FMR mode (solid triangle solid line) and FEX mode (solid diamond solid line) is shown here.\\
		b) The corresponding effective magnetic thickness $t_{eff}$ of the NiFe along the bias direction (empty circle solid line) and against the bias direction (empty square dashed line) for the same range of field cooled temperatures.}
		\end{figure}

Lastly, we discuss measured dependence of bias and pinning on temperature for the 2.5 nm thick IrMn bilayer. This layer was most interesting because it does not show significant bias at room temperature, but does develope bias at lower temperatures. A summary of results is shown in Fig.\ref{fig:temppinning}.  A linear increase in exchange bias below 240K was found from the FMR mode data, and has been reported previously in literature\cite{tsang:2605,cheng:4927-2,PhysRevB.68.214420}. A linear increase in the magnitude of the pinning parameters was found over the same temperature region, with different slopes for $p$ measured parallel and antiparallel to the bias direction.  The behaviour of $t_{eff}$ however reveals similar behavior and slopes for the two field orientations. The interfacial region involved in pinning is determined by the difference between values obtained from parallel and antiparallel orientations. This difference is about 0.5 nm and independent of temperature.

\section{Discussion and Conclusions}
In this paper we have presented results for resonant field shifts due to exchange bias in NiFe/IrMn bilayers. The unidirectional exchange anisotropy was determined from angular resolved resonance experiments. We observed field differences for the lowest order standing spin wave mode that are twice the magnitude of the corresponding difference for the fundamental resonance.  We show that interpretation of these results can be made in terms of pinning effects due to an effective surface exchange anisotropy.  The distortion each spin wave mode experiences due to this pinning is not the same for every mode.  Experimentally this results in different exchange anisotropies observed for FMR and FEX resonances. The assumption of an effective surface anisotropy is possible because resonances of the IrMn are at much higher frequencies than those probed with our coplanar resonance technique, so that the NiFe spin waves are driving the IrMn far off resonance.  Because of this mismatch in frequencies, the effective fields acting on the NiFe spins near the interface are governed by anisotropies induced through exchange coupling to the IrMn, and other dynamics in the antiferromagnet can be safely neglected\cite{PhysRevB.54.4159, PhysRevB.62.6429}. One can understand the pinning simply as a unidirectional anisotropy whose magnitude varies as $\cos (\theta)$, where $\theta$ is the angle of the static field relative to the bias direction.

When calculating the wavevectors of the FMR and FEX modes, deviations from values expected assuming no pinning are found.  Analysing the data this way returns a pinning parameter that charaterises the strength of interface coupling and gives an effective magnetic thickness over which the NiFe film acts as a saturated ferromagnet.
As the structural thickness of the NiFe films are well known, deviations from this value in $t_{eff}$ may arise from the magnetisation close to the interface.  Thus one can also interpret the observed effective thickness as an exchange bias effect that involves a deformation of the magnetization near the interface that reduces the magnetic thickness of the ferromagnet participating in the spin wave resonance. Such a deformation might be possible through either pinning of ferromagnetic spins near the interface, or formation of a twist on the ferromagnet side of the interface. We note that this interpretation is analogous to the effective boundary conditions derived by Guslienko and Slavin for dipolar contributions to resonance in stripes \cite{PHYS.REV.B.72.014463.(2005)}.

We close with two final remarks. First, there exists a difference between exchange bias measurements between FMR and MOKE of at most 30\%.  This is a well known effect\cite{Measure_Bias_MOM_FMR} and is due primarily to FMR being a perturbative measurement of local fields whereas MOKE measurements of hysteresis necessarily involve magnetization processes. Though there has not previously  been an FEX to MOKE comparison, we note that FEX follows the same trend as the FMR data, but with different magnitude as both are perturbative measures of the exchange anisotropy.  Secondly, possible effects associated with field cooling were also sought. As shown above, the 2.5nm IrMn sample has a blocking temperature below room temperature and that it does not experience significant exchange biasing until below 240K.

\section{Acknowledgements}
Support from the Australian Research Council under the Discovery and Australian Postgraduate Award programmes is acknowledged.  Furthermore support from the United Kingdom's Engineering and Physical Sciences Research Council is also acknowledged.



%

\end{document}